\documentclass[twocolumn,prd,nofootinbib]{revtex4}
\usepackage{amsmath}
\usepackage{graphicx}

\begin{document}

\title{Gravitational dynamics in $s+1+1$ dimensions}
\author{L\'{a}szl\'{o} \'{A}. Gergely$^{1}$ and Zolt\'{a}n Kov\'{a}cs$^{2}$}
\affiliation{$^{1}$Departments of Theoretical and Experimental Physics, University of
Szeged, Szeged 6720, D\'{o}m t\'{e}r 9, Hungary\\
$^{2}$Max-Planck-Institut f\"{u}r Astronomie, K\"{o}nigstuhl 17, D-69117
Heidelberg, Germany}

\begin{abstract}
We present the concomitant decomposition of an $\left( s+2\right) $%
-dimensional space-time both with respect to a timelike and a spacelike
direction. The formalism we develop is suited for the study of the initial
value problem and for canonical gravitational dynamics in brane-world
scenarios. The bulk metric is replaced by two sets of variables. The first
set consist of one tensorial (the induced metric $g_{ij}$), one vectorial ($%
M^{i}$) and one scalar ($M$)\ dynamical quantity, all defined on the $s$%
-space. Their time evolutions are related to the second fundamental form
(the extrinsic curvature $K_{ij}$), the normal fundamental form ($\mathcal{K}%
^{i}$) and normal fundamental scalar ($\mathcal{K}$), respectively. The
non-dynamical set of variables is given by the lapse function and the shift
vector, which however has one component less. The missing component is due
to the externally imposed constraint, which states that physical
trajectories are confined to the $\left( s+1\right) $-dimensional brane. The
pair of dynamical variables ($g_{ij}$, $K_{ij}$), well-known from the ADM\
decomposition is supplemented by the pairs ($M^{i}$, $\mathcal{K}^{i}$) and (%
$M$, $\mathcal{K}$) due to the bulk curvature. We give all projections of
the junction condition across the brane and prove that for a perfect fluid
brane neither of the dynamical variables has jump across the brane. Finally
we complete the set of equations needed for gravitational dynamics by
deriving the evolution equations of $K_{ij}$, $\mathcal{K}^{i}$ and $%
\mathcal{K}$ on a brane with arbitrary matter.
\end{abstract}

\maketitle

\section{Introduction}

Recent models motivated by string theory / M theory \cite{ADD}, \cite{RS2}
indicate that gravity may act in more than four non-compact dimensions (the
bulk). Our everyday experiences are preserved by postulating that ordinary
standard-model matter remains confined to the brane, which is a single
spatial time-evolving $3$-surface (a temporal $4$-surface). Remarkably,
gravity is also shown to be localized on the brane in these models \cite{RS2}%
. As energy-momentum is the source for gravity, the expectation is that some
sort of localization of gravity on the brane is quite generic. More exactly,
the Ricci-component of the curvature should be localized through the bulk
Einstein equations. However the Weyl-component of the curvature induced by
black holes in the bulk can give rise to the non-localization of gravity. It
has been known for a while that gravity is not localized on the
Friedmann-Lema\={\i}tre-Robertson-Walker brane embedded in
Schwarzschild-Anti de Sitter bulk, whenever the former has negative
cosmological constant\cite{DS}. A recent example in this sense was found in %
\cite{SCM}, where it is explicitly shown how gravity is delocalized on a
vacuum Einstein brane by the 5-dimensional (5D) horizon of a bulk black
hole. The way to see whether gravity is localized on the brane is a
perturbative one, emerging from the perturbative analysis of the massive
Kaluza-Klein modes of the bulk graviton by perturbations around a Minkowski
brane in anti-de Sitter bulk \cite{RS2} (or Einstein static vacuum brane %
\cite{Einbrane} in Schwarzshild-Anti de Sitter bulk \cite{SCM}).

In order to monitor gravitational dynamics from the brane observer's
viewpoint and to follow the evolution of matter fields on the brane, the
decomposition of bulk quantities and their dynamics with respect to the
brane is necessary. This was done in \cite{SMS}, leading to an effective
Einstein equation on the brane, which contains additional new source terms:
a quadratic expression of the energy-momentum tensor and the ''electric''
part of the bulk Weyl tensor. Brane matter is related to the discontinuity
(across the brane) in the second fundamental form of the brane through the
Lanczos-Sen-Darmois-Israel matching conditions \cite{Lanczos}-\cite{Israel}.
The more generic situation, allowing for a brane embedded asymmetrically and
for matter (non-standard-model fields) in the bulk was presented in \cite%
{Decomp}. There it was proven that the 5D Einstein equations are equivalent
on the brane with the set of the effective Einstein equations (with even
more new source terms, arising from asymmetric embedding and bulk matter),
the Codazzi equation and the twice-contracted Gauss equation. For a recent
review of other issues related to brane-worlds, see \cite{MaartensLR}.

No canonical description of the bulk, similar to the standard Arnowitt-Derser-Misner (ADM) treatment
of the 4D gravity has been given until now. That would be straigthforward if
one simply aims to increase the dimension of space by one. Such a procedure
however would not know about the preferred hypersurface which is the brane.
Occurence of effects like the localization of gravity on the brane would be
difficult to follow. Therefore we propose to develop a formalism which
singles out both the time and the off-brane dimension. Although we primarily
have in mind the $3+1+1$ brane-world model, we would like to keep a more
generic setup for other possible applications, like the $2+1+1$
decomposition of space-time in general relativity. Therefore we develop the
formalism in $s+1+1$ dimensions.

\begin{figure}[tbp]
\vskip-0.5cm \includegraphics[height=6cm]{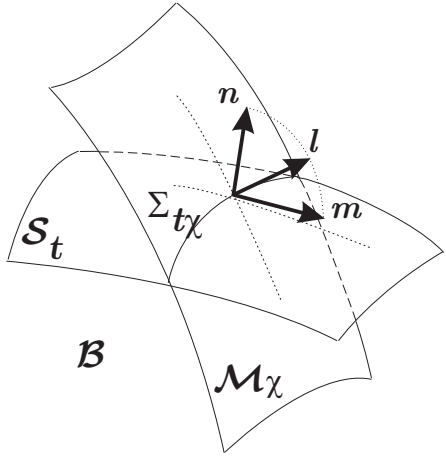}
\caption{The two foliations with unit normal vector fields $n$ and $l$. The
unit vector field $m$ belongs to the tangent space spanned by $n$ and $l$
and it is perpendicular to $n$.}
\label{Fig1}
\end{figure}

We suppose that the full non-compact space-time $\mathcal{B}$ can be
foliated by a family of $(s+1)$-dimensional space-like leaves $S_{t}$,
characterized by the unit normal vector field $n$. The projection of this
foliation onto the observable $(s+1)$-dimensional space-time $\mathcal{M}$
(the brane) gives the usual $s+1$ decomposition of $\mathcal{M}$ into the
foliation $\Sigma _{t}$, whenever $n\left( x\in \mathcal{M}\right) \in T%
\mathcal{M}.$ Let us denote the unit normal\ vector field to $\mathcal{M}$
by $l$. We also suppose that $\mathcal{B}$ can be foliated by a family of $%
(s+1)$-surfaces $\mathcal{M}_{\chi }$ (with $\mathcal{M}_{0}=\mathcal{M}$),
or at least we are able to extend the field $l$ in a suitable manner to some
neighborhood of $\mathcal{M}$. We denote by $m$ the unit normal to $n$ in
the tangent plane spanned by $n$ and $l$. The intersection of the leaves $%
S_{t}$ and $\mathcal{M}_{\chi }$ represent spatial $s$-surfaces $\Sigma
_{t\chi }$, from among which $\Sigma _{t}=\Sigma _{t0}$ (Fig. \ref{Fig1}).

In Sec.II we define temporal and off-brane evolution vectors and we
decompose them in an orthonormal basis adapted to the foliation $S_{t}$. The 
$\left( s+2\right) $-dimensional metric is replaced by a convenient set of
dynamical variables $\left( g_{ij}\text{, }M^{i}\text{, }M\right) $,
together with the lapse and shifts of $n$. We show that one shift component, 
$\mathcal{N}$ obeys a constraint due to the Frobenius theorem. This
constraint can be traced back to the fact that the trajectories of standard
model particles are confined to a hypersurface. In other words, it is a
consequence of the very existence of the brane. The simplest way to fulfill
this constraint is to choose a vanishing $\mathcal{N}$. This in turn is
equivalent with choosing the two foliations perpendicularly, $m^{a}=l^{a}$.
Then we introduce the second fundamental forms of the $S_{t}$, $\mathcal{M}%
_{\chi }$ and $\Sigma _{t\chi }$ hypersurfaces and establish their
interconnections. This is done via the normals $n$ and $l$, their
''accelerations'', the normal fundamental forms and normal fundamental
scalars, all to be defined in Sec. II. Some of the more technical details
needed for the results of Sec.II are derived in Appendices A, B and C.

We establish in Sec.III the relation between time-derivatives of the
dynamical data and various projections of extrinsic curvatures $\left( K_{ij}%
\text{, }\mathcal{K}^{i}\text{, }\mathcal{K}\right) $. In Sec.IV we write
the junction conditions, in particular the Lanczos equation across a brane
containing arbitrary matter in terms of these extrinsic curvature
projections.

Sec.V and Appendix D contain the decompositions with respect to $S_{t}$, $%
\mathcal{M}_{\chi }$ and $\Sigma _{t\chi }$ of the intrinsic curvatures. We
give the decomposition of the connections, Riemann, Ricci and Einstein
tensors. Among these we find the Raychaudhuri, Codazzi and Gauss equations.
The evolution equations for the set $\left( K_{ij}\text{, }\mathcal{K}^{i}%
\text{, }\mathcal{K}\right) $ are then readily deduced in Sec.VI. We give
them explicitly for a bulk containing nothing but a negative cosmological
constant, but for a generic brane. With this we complete the task of giving
all gravitational evolution equations in terms of one tensorial, one
vectorial and one scalar pair of dynamical quantities.

Sec.VII contains the Concluding Remarks. Here we compare our formalism
specified for $s=2$ with previous decompositions of space-time in general
relativity.

\textbf{Notation.} A tilde and a hat distinguish the quantities defined on $%
\mathcal{B}$ and $S_{t}$, respectively. Quantities belonging to $\mathcal{M}%
_{\chi }$ posess a distinctive dimension-carrying index while those defined
on $\Sigma _{t\chi }$ have no special distinctive mark. For example, the
metric 2-forms on $\mathcal{B}$, $S_{t}$, $\mathcal{M}_{\chi }$ and $\Sigma
_{t\chi }$ are denoted $\widetilde{g}$, $\widehat{g},$ ${^{(s+1)\!}g}$ and$\
g$, respectively, while the corresponding metric-compatible connections are $%
\widetilde{\nabla }$, $\widehat{D}$, $^{(s+1)\!\!\!}D$ and $D$. Then 
\begin{eqnarray}
\widehat{g}_{c_{1}...c_{r}b_{1}...b_{s}}^{a_{1}...a_{r}d_{1}...d_{s}} &=&%
\widehat{g}_{c_{1}}^{a_{1}}{}...\widehat{g}_{c_{r}}^{a_{r}}{}\widehat{g}%
_{b_{1}}^{d_{1}}{}...\widehat{g}_{b_{s}}^{d_{s}}{}\ ,  \notag \\
^{\left( s+1\right)
}\!g_{c_{1}...c_{r}b_{1}...b_{s}}^{a_{1}...a_{r}d_{1}...d_{s}} &=&{%
^{(s+1)\!\!}}g_{c_{1}}^{a_{1}}{}...{^{(s+1)\!\!}g}_{c_{r}}^{a_{r}}{}{%
^{(s+1)\!\!}g}_{b_{1}}^{d_{1}}{}...{^{(s+1)\!\!}g}_{b_{s}}^{d_{s}}{}\ , 
\notag \\
g_{c_{1}...c_{r}b_{1}...b_{s}}^{a_{1}...a_{r}d_{1}...d_{s}}
&=&g_{c_{1}}^{a_{1}}{}...g_{c_{r}}^{a_{r}}{}g_{b_{1}}^{d_{1}}{}...g_{b_{s}}^{d_{s}}{}\ ,
\end{eqnarray}%
project any tensor $\widetilde{T}_{b_{1}...b_{s}}^{a_{1}...a_{r}}$ on $%
\mathcal{B}$ to $S_{t}$, $\mathcal{M}_{\chi }$ and $\Sigma _{t\chi }$,
respectively.

Latin indices represent abstract indices running from $0$ to $(s+1)$. Greek
and bold latin indices, running from $0$ to $(s+1)$ and from $1$ to $s$,
respectively either count some specific basis vectors or they denote
tensorial components in these bases. Vector fields in Lie-derivatives are
represented by boldface characters. For example $\widetilde{\mathcal{L}}_{%
\mathbf{V}}T$ denotes the Lie derivative on $\mathcal{B}$ along the integral
lines of the vector field $V^{a}$.

\section{Fundamental forms}

\subsection{First fundamental forms of $\Sigma _{t\protect\chi }$ , $S_{t}$
and $\mathcal{M}_{\protect\chi }$}

\begin{figure}[tbp]
\vskip-0.5cm \includegraphics[height=6cm]{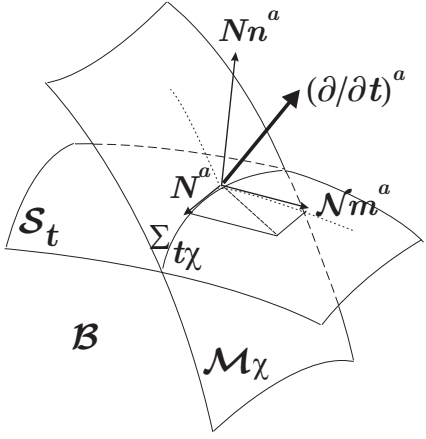}
\caption{Decomposition of the time-evolution vector $\partial /\partial t$,
when the two foliations are not perpendicular.}
\label{Fig2}
\end{figure}

By a careful study of the two foliations in Appendix A we arrive to the
conclusion that temporal and off-brane evolutions happen along vector fields
given as: 
\begin{eqnarray}
\left( \frac{\partial }{\partial t}\right) ^{a} &=&Nn^{a}+N^{a}+\mathcal{N}%
m^{a}\ ,  \label{ddt0} \\
\left( \frac{\partial }{\partial \chi }\right) ^{a} &=&M^{a}+Mm^{a}\ .
\label{ddchi0}
\end{eqnarray}%
In the above formulae $N^{a}$ and $N$ have the well-known interpretation
from the decomposition of the $3+1$ dimensional space-time as shift vector
and lapse function. The quantity $\mathcal{N}$ is the component of the shift
in the off-brane direction (Fig. \ref{Fig2}). 
Finally the vector $M^{a}$ and the scalar $M$ are quantities representing
the off-brane sector of gravity, all defined on $\Sigma _{t\chi }$ (Fig. \ref%
{Fig3}). 
\begin{figure}[tbp]
\vskip-0.5cm \includegraphics[height=6cm]{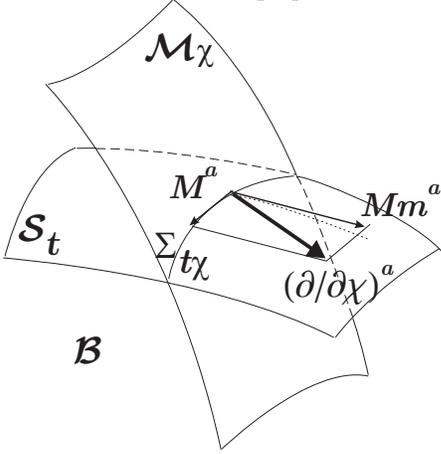}
\caption{Decomposition of the off-brane evolution vector $\partial /\partial 
\protect\chi $ for non-perpendicular foliations.}
\label{Fig3}
\end{figure}

The $\left( s+2\right) $-dimensional metric is%
\begin{equation}
\widetilde{g}_{ab}=g_{ab}+m_{a}m_{b}-n_{a}n_{b}\ ,  \label{tildeg0}
\end{equation}%
$g_{ab}$ being the induced metric of $\Sigma _{t\chi }$. In Appendix B we
proove a simple relationship between the corresponding determinants: 
\begin{equation}
\sqrt{-\widetilde{g}}=NM\sqrt{g}\ .  \label{gdet}
\end{equation}%
The induced metrics on $S_{t}$ and on $\mathcal{M}_{\chi }$ are
respectively: 
\begin{eqnarray}
\widehat{g}_{ab} &=&\widetilde{g}_{ab}+n_{a}n_{b}=g_{ab}+m_{a}m_{b}\ ,
\label{ghat} \\
{^{(s+1)\!\!}}g_{ab} &=&\widetilde{g}_{ab}-m_{a}m_{b}=g_{ab}-n_{a}n_{b}\ .
\label{4g}
\end{eqnarray}%
They obey $\widehat{g}_{ab}n^{b}={^{(s+1)\!}g}_{ab}l^{b}=0$.

\begin{figure}[tbp]
\vskip-0.5cm \includegraphics[height=6cm]{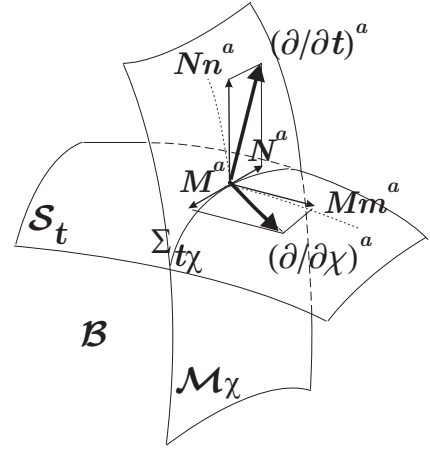}
\caption{Decomposition of the temporal and off-brane evolution vectors $%
\partial /\partial t$ and $\partial /\partial \protect\chi $ for
perpendicular foliations.}
\label{Fig4}
\end{figure}

A simple counting shows that the $\left( s+2\right) \left( s+3\right) /2$
components of the $(s+2)$-metric $\widetilde{g}_{\mathbf{\alpha \beta }}$
can be replaced by the equivalent set $\{g_{ab},M^{a},M,N^{a},\mathcal{N}%
,N\} $, for which the restrictions $%
g_{ab}n^{a}=g_{ab}m^{a}=M^{a}n_{a}=M^{a}m_{a}=N^{a}n_{a}=N^{a}m_{a}=0$
apply. In Appendix C we however prove from the Frobenius theorem the
constraint (\ref{Frobenius}) on $\mathcal{N}$ and $M$. This seems to reduce
the number of variables, still the information is not lost. Although the
number of gravitational variables is reduced by one, we gain the information
that the evolution of standard-model fields is constrained to a
hypersurface. The easiest way to satisfy this constraint is to choose 
\begin{equation}
\mathcal{N}=0\ ,
\end{equation}%
which means perpendicular foliations (Fig. \ref{Fig4}). 
We will follow this choice, which leads to the identification 
\begin{equation}
l^{a}=m^{a}
\end{equation}%
in the rest of the paper (excepting Appendices A, B and C, the results of
which are indispensable in deriving the constraint (\ref{Frobenius})).

\subsection{Second fundamental forms of $\Sigma _{t\protect\chi }$ , $S_{t}$
and $\mathcal{M}_{\protect\chi }$}

We introduce the $\Sigma _{t\chi }$, $S_{t}$ and $\mathcal{M}_{\chi }$%
-projections of the covariant derivative of an arbitrary tensor $\widetilde{T%
}_{b_{1}...b_{s}}^{a_{1}...a_{r}}$ defined on $\mathcal{B}$ as: 
\begin{eqnarray}
D_{a}\widetilde{T}_{b_{1}...b_{s}}^{a_{1}...a_{r}}
&=&g_{ac_{1}...c_{r}b_{1}...b_{s}}^{ca_{1}...a_{r}d_{1}...d_{s}}\widetilde{%
\nabla }_{c}\widetilde{T}_{d_{1}...d_{s}}^{c_{1}...c_{r}}\ ,  \notag \\
\widehat{D}_{a}\widetilde{T}_{b_{1}...b_{s}}^{a_{1}...a_{r}} &=&\widehat{g}%
_{ac_{1}...c_{r}b_{1}...b_{s}}^{ca_{1}...a_{r}d_{1}...d_{s}}\widetilde{%
\nabla }_{c}\widetilde{T}_{d_{1}...d_{s}}^{c_{1}...c_{r}}\ ,  \notag \\
^{\left( s+1\right) }D_{a}\widetilde{T}_{b_{1}...b_{s}}^{a_{1}...a_{r}} &=&{%
^{(s+1)\!\!}}g_{ac_{1}...c_{r}b_{1}...b_{s}}^{ca_{1}...a_{r}d_{1}...d_{s}}%
\widetilde{\nabla }_{c}\widetilde{T}_{d_{1}...d_{s}}^{c_{1}...c_{r}}\ .
\label{projcovderiv}
\end{eqnarray}%
When $\widetilde{T}_{b_{1}...b_{s}}^{a_{1}...a_{r}}$ coincides with its
projection to $\Sigma _{t\chi }$, $S_{t}$ or $\mathcal{M}_{\chi }$, the
expressions (\ref{projcovderiv}) are the covariant derivatives in $\Sigma
_{t\chi }$, $S_{t}$ or $\mathcal{M}_{\chi }$, respectively (they annihilate
the corresponding metrics). Despite the notation, projections of derivatives
of tensors not lying on $\Sigma _{t\chi }$, $S_{t}$ and $\mathcal{M}_{\chi }$
are not derivatives, as they fail to obey the Leibniz rule. For example: $%
D_{a}(Ml^{b})=MD_{a}l^{b}.$

As there are two normals to $\Sigma _{t\chi }$, we can define two kinds of
extrinsic curvatures:

\begin{eqnarray}
K_{ab} &=&D_{a}n_{b}=g_{ab}^{cd}{}\widetilde{\nabla }_{c}n_{d}\ ,  \notag \\
L_{ab} &=&D_{a}l_{b}=g_{ab}^{cd}{}\widetilde{\nabla }_{c}l_{d}\ .
\label{extr3}
\end{eqnarray}%
We denote their traces by $K$ and $L$. It is immediate to see the symmetry
of these extrinsic curvatures by noting that $n_{b}=-N \widetilde{\nabla }%
_{b}t$ and $l_{b}=M \widetilde{\nabla }_{b}\chi $ (see Eqs. (\ref{fdualbasis}%
)).

It is also useful to introduce the extrinsic curvatures of $S_{t}$ :

\begin{equation}
\widehat{K}_{ab}=\widehat{D}_{a}n_{b}=\widehat{g}_{ab}^{cd}{}\widetilde{%
\nabla }_{c}n_{d}=\widetilde{\nabla }_{a}n_{b}+n_{a}n^{c}\widetilde{\nabla }%
_{c}n_{b.}\ .  \label{hatKab}
\end{equation}%
As $n^{a}$ and $l^{a}$ are hypersurface-orthogonal, all extrinsic curvatures
defined above are symmetric.

An $\left( s+1\right) +1$ decomposition would be the generalization to
arbitrary dimension of the ADM decomposition of gravity, in which the
extrinsic curvature $\widehat{K}^{ab}$ takes the role of the canonical
momenta associated to $\widehat{g}_{ab}$. Instead we would like a formalism
which singles out both directions $n$ and $l$. By identifying $\widehat{g}%
_{ab}\equiv \{g_{ab},M^{a},M,\mathcal{N},N^{a}\}$ (see Eqs. (\ref{tildeg}), (%
\ref{ghat}) and (\ref{fdualbasis})), it is immediate to choose the set $%
\{g_{ab},M^{a},M\}$ as canonical coordinates on $S_{t}$ and to search for
the canonical momenta among the various projections of $\widehat{K}_{ab}$:

\begin{equation}
\widehat{K}_{ab}=K_{ab}+2l_{(a}\mathcal{K}_{b)}+l_{a}l_{b}\mathcal{K\ }.
\label{Khatdecomp}
\end{equation}%
The quantities 
\begin{eqnarray}
\mathcal{K}_{a} &=&{^{(s+1)\!\!}}g_{a}^{c}l^{d}\widehat{K}_{cd}\ ,  \notag \\
\mathcal{K} &=&l^{c}l^{d}\widehat{K}_{cd}\   \label{calK}
\end{eqnarray}%
represent off-brane projections of $\widehat{K}_{ab}$. $\mathcal{K}^{i}$ is
the normal fundamental form, introduced in \cite{Spivak} and we call $%
\mathcal{K}$ the normal fundamental scalar. The extrinsic curvature $K_{ab}$
defined in (\ref{extr3}) can be shown to be the projection of $\widehat{K}%
_{ab}$ to $\Sigma _{t\chi }$: 
\begin{equation}
K_{ab}={^{(s+1)\!\!}}g_{ab}^{cd}\ \widehat{K}_{cd}  \label{KhatK}
\end{equation}%
from the $g_{a}^{c}{}={^{(s+1)\!\!}}g_{d}^{c}\widehat{g}_{a}^{d}{}$ property
of the projectors. We remark that $l^{a}K_{ab}=0$ and $l^{a}\mathcal{K}%
_{a}=0 $ hold, thus $K_{ab}$ and $\mathcal{K}_{a}$ are tensors defined on $%
\Sigma _{t\chi }$.

In terms of the $\left( s+2\right) $-connection $\widetilde{\nabla }$, the
projections $\mathcal{K}_{a}$ and $\mathcal{K}$ are expressed as: 
\begin{eqnarray}
\mathcal{K}_{a} &=&g_{a}^{b}{}l^{c}\widetilde{\nabla }%
_{b}n_{c}=g_{a}^{b}{}l^{c}\widetilde{\nabla }_{c}n_{b}\ ,  \notag \\
\mathcal{K} &=&l^{a}l^{b}\widetilde{\nabla }_{a}n_{b}=l^{a}\widetilde{%
\mathcal{L}}_{\mathbf{n}}l_{a}\ .  \label{calK1}
\end{eqnarray}%
The second expression for $\mathcal{K}_{a}$ follows from the symmetry of $%
\widehat{K}_{ab}$ in Eq. (\ref{hatKab}).

Alternatively, keeping in mind the bulk-brane scenario, where a
decomposition with respect to the normal $l^{a}$ is frequently desirable %
\cite{SMS}, we introduce the extrinsic curvature of the space-time leave $%
\mathcal{M}_{\chi }$ : 
\begin{equation}
{^{(s+1)\!\!}}L_{ab}={^{(s+1)\!\!}}D_{a}l_{b}={^{(s+1)\!\!}g}_{ab}^{cd}{}%
\widetilde{\nabla }_{c}l_{d}=\widetilde{\nabla }_{a}l_{b}-l_{a}l^{c}%
\widetilde{\nabla }_{c}l_{b}\ ,  \label{4Lab}
\end{equation}%
and we decompose it with respect to the $\Sigma _{t\chi }$ foliation as

\begin{equation}
{^{(s+1)\!\!}}L_{ab}=L_{ab}+2n_{(a}\mathcal{L}_{b)}+n_{a}n_{b}\mathcal{L\ }.
\label{L4abdecomp}
\end{equation}%
Here the quantities

\begin{eqnarray}
\mathcal{L}_{a} &=&-\widehat{g}_{a}^{b}n^{c}{^{(s+1)\!\!}}L_{bc}\ ,  \notag
\\
\mathcal{L} &=&n^{a}n^{b}{^{(s+1)\!\!}}L_{ab}  \label{4L1}
\end{eqnarray}%
represent timelike projections of ${^{(s+1)\!\!}}L_{ab}$ and the previously
introduced extrinsic curvature $L_{ab}$ is nothing but the projection of ${%
^{(s+1)\!\!}}L_{ab}$ to $\Sigma _{t\chi }$: 
\begin{equation}
L_{ab}=\widehat{g}_{ab}^{cd}\ {^{(s+1)\!\!}}L_{cd}\ .  \label{L4L}
\end{equation}%
As $n^{a}L_{ab}=0$ and $n^{a}\mathcal{L}_{a}=0$ indicate, both $L_{ab}$ and $%
\mathcal{L}_{a}$ are tensors defined on $\Sigma _{t\chi }$.

The projections $\mathcal{L}_{a}$ and $\mathcal{L}$ can be equally expressed
as:

\begin{eqnarray}
\mathcal{L}_{a} &=&-g_{a}^{b}{}n^{c}\widetilde{\nabla }%
_{b}l_{c}=-g_{a}^{b}{}n^{c}\widetilde{\nabla }_{c}l_{b}=\mathcal{K}_{a}\ , 
\notag \\
\mathcal{L} &=&n^{a}n^{b}\widetilde{\nabla }_{a}l_{b}=n^{a}\widetilde{%
\mathcal{L}}_{\mathbf{l}}n_{a}\ .  \label{4L2}
\end{eqnarray}%
Thus $\mathcal{L}_{a}$ will be replaced in all forthcoming computations by $%
\mathcal{K}_{a}$.

We have just completed the task to characterize both the extrinsic
curvatures of $S_{t}$ and of $\mathcal{M}_{\chi }$ in terms of quantities
defined on $\Sigma _{t\chi }$ alone.

Finally we define the curvatures $\alpha ^{a}$ and $\lambda ^{a}$ of the
congruences $n^{a}$ and $l^{a}$ 
\begin{eqnarray}
\alpha ^{b} &=&n^{c}\widetilde{\nabla }_{c}n^{b}=g_{d}^{b}{}n^{c}\widetilde{%
\nabla }_{c}n^{d}-\mathcal{L}l^{b}\ ,  \label{acc} \\
\lambda ^{b} &=&l^{c}\widetilde{\nabla }_{c}l^{b}=g_{d}^{b}{}l^{c}\widetilde{%
\nabla }_{c}l^{d}+\mathcal{K}n^{b}\ .  \label{lambda}
\end{eqnarray}%
The second pair of expressions for each of the above curvatures are nothing
but their $s+1+1$ decomposition. Physically, the curvature $\alpha ^{a}$ is
the nongravitational acceleration of observers with velocity $n^{a}$.
Detailed expressions for these curvatures are deduced in Appendix C.

Let us note, that the sets of Eqs. (\ref{hatKab})-(\ref{Khatdecomp}) and (%
\ref{4Lab})-(\ref{L4abdecomp}) allow for the following decompositions of the
covariant derivatives of the normals: 
\begin{eqnarray}
\widetilde{\nabla }_{a}n_{b} &=&K_{ab}+2\mathcal{K}_{(a}l_{b)}+l_{a}l_{b}%
\mathcal{K}-n_{a}\alpha _{b}\ ,  \notag \\
\widetilde{\nabla }_{a}l_{b} &=&L_{ab}+2\mathcal{K}_{(a}n_{b)}+n_{a}n_{b}%
\mathcal{L}+l_{a}\lambda _{b}\ .  \label{normderivdecomp}
\end{eqnarray}%
From here we find simple expressions for their covariant divergences: 
\begin{eqnarray}
\widetilde{\nabla }_{a}n^{a} &=&\widehat{K}=K+\mathcal{K\ },  \notag \\
\widetilde{\nabla }_{a}l^{a} &=&\widehat{L}=L-\mathcal{L\ }.
\label{covardivs}
\end{eqnarray}%
We also derive the useful relation 
\begin{equation}
\widetilde{\nabla }_{a}n^{b}\widetilde{\nabla }_{b}n^{a}=\widehat{K}_{ab}%
\widehat{K}^{ab}=K_{ab}K^{ab}+2\mathcal{K}_{a}\mathcal{K}^{a}+\mathcal{K}%
^{2}\ .
\end{equation}

\section{Time-derivatives of $g_{ab}$, $M^{a}$ and $M$}

In this section we establish the relations among the time derivatives of the
dynamical data $\{g_{ab},M^{a},M\}$ and extrinsic curvatures $\{K_{ab},%
\mathcal{K}_{a},\mathcal{K}\}$ and $\{L_{ab},\mathcal{L}_{a},\mathcal{L}\}$.

We proceed as follows. First we define suitable derivatives of tensors from $%
S_{t}$ and $\mathcal{M}_{\chi }$ as the respective projections of the
Lie-derivative in $\mathcal{B}$ along an arbitrary vector flow $V^{a}$: 
\begin{eqnarray}
\widehat{\mathcal{L}}_{\mathbf{V}}\widehat{T}%
_{b_{1}...b_{s}}^{a_{1}...a_{r}} &=&\widehat{g}%
_{c_{1}...c_{r}b_{1}...b_{s}}^{a_{1}...a_{r}d_{1}...d_{s}}\widetilde{%
\mathcal{L}}_{\mathbf{V}}\widehat{T}_{d_{1}...d_{s}}^{c_{1}...c_{r}}\ ,
\label{Lieprojhat} \\
{^{(s+1)\!\!}}\mathcal{L}_{\mathbf{V}}{^{(s+1)\!\!}}%
T_{b_{1}...b_{s}}^{a_{1}...a_{r}} &=&{^{(s+1)\!\!}}%
g_{c_{1}...c_{r}b_{1}...b_{s}}^{a_{1}...a_{r}d_{1}...d_{s}}  \notag \\
&&\times \widetilde{\mathcal{L}}_{\mathbf{V}}\ {^{(s+1)\!\!}}%
T_{d_{1}...d_{s}}^{c_{1}...c_{r}}\ .  \label{Lieproj4}
\end{eqnarray}%
When $V^{a}$ belongs to $S_{t}$ ($\mathcal{M}_{\chi }$), the derivative $%
\widehat{\mathcal{L}}_{\mathbf{V}}$ (${^{(s+1)\!\!}}\mathcal{L}_{\mathbf{V}}$%
) is the Lie-derivative in $S_{t}$ ($\mathcal{M}_{\chi }$).

When $\mathbf{V}$ is $\partial /\partial t$ or $\partial /\partial \chi $,
the derivatives decouple as: 
\begin{eqnarray}
\widehat{\mathcal{L}}_{\frac{\mathbf{\partial }}{\mathbf{\partial t}}}%
\widehat{T}_{b_{1}...b_{s}}^{a_{1}...a_{r}} &=&N\widehat{\mathcal{L}}_{%
\mathbf{n}}\widehat{T}_{b_{1}...b_{s}}^{a_{1}...a_{r}}+\widehat{\mathcal{L}}%
_{\mathbf{N}}\widehat{T}_{b_{1}...b_{s}}^{a_{1}...a_{r}}\ ,
\label{Lieprojthat} \\
{^{(s+1)\!\!}}\mathcal{L}_{\frac{\mathbf{\partial }}{\mathbf{\partial \chi }}%
}{^{(s+1)\!\!}}T_{b_{1}...b_{s}}^{a_{1}...a_{r}} &=&M{^{(s+1)\!\!}}\mathcal{L%
}_{\mathbf{l}}{^{(s+1)\!\!}}T_{b_{1}...b_{s}}^{a_{1}...a_{r}}  \notag \\
&&+{^{(s+1)\!\!}}\mathcal{L}_{\mathbf{M}}{^{(s+1)\!\!}}%
T_{b_{1}...b_{s}}^{a_{1}...a_{r}}\ .  \label{Lieprojchi4}
\end{eqnarray}%
In particular, for the metrics induced on $S_{t}$ and $\mathcal{M}_{\chi }$
we find: 
\begin{eqnarray}
\widehat{\mathcal{L}}_{\frac{\mathbf{\partial }}{\mathbf{\partial t}}}%
\widehat{g}_{ab} &=&2N\widehat{K}_{ab}+2\widehat{D}_{(a}N_{b)}\ ,
\label{ghatt} \\
{^{(s+1)\!\!}}\mathcal{L}_{\frac{\mathbf{\partial }}{\mathbf{\partial t}}}{%
^{(s+1)\!\!}g}_{ab} &=&2M{^{(s+1)\!\!}}L_{ab}+2{^{(s+1)\!\!}}D_{(a}M_{b)}\ .
\label{g4chi}
\end{eqnarray}%
We have used that the extrinsic curvatures (\ref{hatKab}) and (\ref{4Lab})
are expressible as 
\begin{eqnarray}
\widehat{K}_{ab} &=&\frac{1}{2}\widehat{\mathcal{L}}_{\mathbf{n}}\widehat{g}%
_{ab}=\frac{1}{2}\widetilde{\mathcal{L}}_{\mathbf{n}}\widehat{g}_{ab}\ , 
\notag \\
{^{(s+1)\!\!}}L_{ab} &=&\frac{1}{2}{^{(s+1)\!\!}}\mathcal{L}_{\mathbf{l}}{%
^{(s+1)\!\!}g}{}_{ab}=\frac{1}{2}\widetilde{\mathcal{L}}_{\mathbf{l}}{%
^{(s+1)\!\!}g}{}_{ab}\ .
\end{eqnarray}

Next we define a projected derivative of a tensor taken from $\Sigma _{t\chi
}\ $\footnote{%
The expressions (\ref{Lieprojhat}), (\ref{Lieproj4}) and (\ref{Lieproj}),
introduced for tensors projected to $S_{t}$, $\mathcal{M}_{\chi }$ or $%
\Sigma _{t\chi }$, when generalized to arbitrary tensors on $\mathcal{B}$
could fail to obey the Leibniz rule. For example: $\mathcal{L}_{\mathbf{V}%
}(Ml^{b})=M\mathcal{L}_{\mathbf{V}}l^{b}$ .} 
\begin{equation}
\mathcal{L}_{\mathbf{V}%
}T_{b_{1}...b_{s}}^{a_{1}...a_{r}}=g_{c_{1}...c_{r}b_{1}...b_{s}}^{a_{1}...a_{r}d_{1}...d_{s}}%
\widetilde{\mathcal{L}}_{\mathbf{V}}T_{d_{1}...d_{s}}^{c_{1}...c_{r}}\ ,
\label{Lieproj}
\end{equation}%
in terms of which time- and off-brane derivatives can be defined as the
projections to $\Sigma _{t,\chi }$ of the Lie-derivatives $\widetilde{%
\mathcal{L}}$ taken along $\partial /\partial t$ and $\partial /\partial
\chi $ directions, respectively: 
\begin{eqnarray}
\frac{\partial }{\partial t}T_{b_{1}...b_{s}}^{a_{1}...a_{r}} &=&\mathcal{L}%
_{\frac{\mathbf{\partial }}{\mathbf{\partial t}}%
}T_{b_{1}...b_{s}}^{a_{1}...a_{r}}  \notag \\
&=&N\mathcal{L}_{\mathbf{n}}T_{b_{1}...b_{s}}^{a_{1}...a_{r}}+\mathcal{L}_{%
\mathbf{N}}T_{b_{1}...b_{s}}^{a_{1}...a_{r}}\ ,  \label{derivt} \\
\frac{\partial }{\partial \chi }T_{b_{1}...b_{s}}^{a_{1}...a_{r}} &=&%
\mathcal{L}_{\frac{\mathbf{\partial }}{\mathbf{\partial \chi }}%
}T_{b_{1}...b_{s}}^{a_{1}...a_{r}}  \notag \\
&=&M\mathcal{L}_{\mathbf{l}}T_{b_{1}...b_{s}}^{a_{1}...a_{r}}+\mathcal{L}_{%
\mathbf{M}}T_{b_{1}...b_{s}}^{a_{1}...a_{r}}\ .  \label{derivchi}
\end{eqnarray}%
It is not difficult to show that despite of being \textit{projected }%
Lie-derivatives, they become partial derivatives in any adapted coordinate
system. Thus for any $T_{b_{1}...b_{s}}^{a_{1}...a_{r}}$ defined on $\Sigma
_{t,\chi }$ the property 
\begin{equation}
\frac{\partial }{\partial t}\frac{\partial }{\partial \chi }%
T_{b_{1}...b_{s}}^{a_{1}...a_{r}}=\frac{\partial }{\partial \chi }\frac{%
\partial }{\partial t}T_{b_{1}...b_{s}}^{a_{1}...a_{r}}
\end{equation}%
holds. For the $s$-metric the formulae (\ref{derivt})-(\ref{derivchi})
reduce to: 
\begin{eqnarray}
\frac{\partial }{\partial t}g_{ab} &=&2NK_{ab}+2D_{(a}N_{b)}\ ,
\label{gtextr} \\
\frac{\partial }{\partial \chi }g_{ab} &=&\ 2ML_{ab}+2D_{(a}M_{b)}\ ,
\label{gchiextr}
\end{eqnarray}%
where we have used that the extrinsic curvatures of $\Sigma _{t\chi }$ can
be expressed as: 
\begin{eqnarray}
K_{ab} &=&\frac{1}{2}\mathcal{L}_{\mathbf{n}}g_{cd}\ ,  \notag \\
L_{ab} &=&\frac{1}{2}\mathcal{L}_{\mathbf{l}}g_{cd}\ .  \label{extrLie}
\end{eqnarray}%
Eqs. (\ref{gtextr}) and (\ref{gchiextr}), when inverted with respect to the
extrinsic curvatures, read: 
\begin{eqnarray}
\ K_{ab} &=&\frac{1}{2N}\left( \frac{\partial }{\partial t}%
g_{ab}-2D_{(a}N_{b)}\right) \ ,  \label{Kab} \\
L_{ab} &=&\frac{1}{2M}\left( \frac{\partial }{\partial \chi }%
g_{ab}-2D_{(a}M_{b)}\right) \ .  \label{Lab}
\end{eqnarray}

The comparison of the rhs-s from Eqs. (\ref{acc}) and (\ref{accdet}), also
from (\ref{lambda}) and (\ref{lambdadet}) leads to simple expressions for
the curvatures $\alpha ^{b}$ and $\lambda ^{b}$: 
\begin{eqnarray}
\alpha ^{b} &=&D^{b}\left( \ln N\right) -\mathcal{L}l^{b}\ ,
\label{accalpha} \\
\lambda ^{b} &=&-D^{b}\left( \ln M\right) +\mathcal{K}n^{b}\ ,
\label{acclambda}
\end{eqnarray}%
with $\mathcal{L}$ and $\mathcal{K}$ given by 
\begin{eqnarray}
\mathcal{L} &=&-\frac{1}{MN}\left( \frac{\partial }{\partial \chi }%
N-M^{a}D_{a}N\right) \ ,  \label{Lcal} \\
\mathcal{K} &=&\frac{1}{MN}\left( \frac{\partial }{\partial t}%
M-N^{a}D_{a}M\right) \ .  \label{Kcal}
\end{eqnarray}
Equivalently: 
\begin{equation}
\mathcal{K}=\frac{1}{M}\mathcal{L}_{\mathbf{n}}M \ , \quad \mathcal{L}=-%
\frac{1}{N}\mathcal{L}_{\mathbf{l}}N\ .
\end{equation}

\bigskip Finally the difference of the contractions of Eqs. (\ref%
{normderivdecomp}) with $l^{a}$ and $n^{a}$ respectively leads to $2\mathcal{%
K}^{a}=\left( [l,n]^{a}-\mathcal{K}l^{a}-\mathcal{L}n^{a}\right) $.
Inserting the commutator (\ref{nl}) and keeping in mind that $\partial
M^{a}/\partial t=\left( F_{\mathbf{i}}\right) ^{a}\partial M^{\mathbf{i}%
}/\partial t$ and $\partial N^{a}/\partial \chi =\left( F_{\mathbf{i}%
}\right) ^{a}\partial N^{\mathbf{i}}/\partial \chi $ hold in the chosen
coordinate basis $\left\{ e_{\beta }\right\} $, we get 
\begin{equation}
\mathcal{K}^{a}=\frac{1}{2MN}\left[ \frac{\partial }{\partial t}M^{a}-\frac{%
\partial }{\partial \chi }N^{a}+M^{b}D_{b}N^{a}-N^{b}D_{b}M^{a}\right] \ .
\label{Kacal}
\end{equation}
In terms of Lie-derivatives this gives 
\begin{eqnarray}
\mathcal{K}^{a}&=&\frac{1}{2M}\mathcal{L}_{\mathbf{n}}M^{a} -\frac{1}{2N}%
\mathcal{L}_{\mathbf{l}}N^{a} \notag \\
&&-\frac{1}{2NM}(M^{c}K_{c}^{a}-N^{c}L_{c}^{a}) \ .
\end{eqnarray}

Summarizing, Eqs. (\ref{Kab}), (\ref{Kcal}) and (\ref{Kacal}) express the
projections of the extrinsic curvature $\widehat{K}_{ab}$ in terms of the
time derivatives $\partial g_{ab}/\partial t,\ \partial M^{a}/\partial t,\
\partial M/\partial t$ and other quantities from $\Sigma _{t\chi }$.
Concerning the extrinsic curvature of $\mathcal{M}_{\chi }$, Eqs. (\ref{Lab}%
) and (\ref{Lcal}) convinces us that $L_{ab\text{ }}$and $\ \mathcal{L}$ are
those projections of ${^{(s+1)\!\!}}L_{ab}$, which have nothing to do with
time evolution, while the projection $\mathcal{L}^{a}=\mathcal{K}^{a}$ is
dynamical.

\section{The Lanczos equation}

Whenever the hypersurface $\mathcal{M}$ has a distributional energy momentum 
$\tau _{ab}$, the extrinsic curvature ${^{(s+1)\!\!}}L_{ab}$ will have a jump%
\begin{equation}
\Delta {^{(s+1)\!\!}}L_{ab}=-\widetilde{\kappa }^{2}\left( \tau _{ab}-\frac{1%
}{s}{^{(s+1)\!\!}}g_{ab}\tau \right) \ .  \label{Lanczos1}
\end{equation}%
This is particularly interesting in the brane-world scenario. Then $\tau
_{ab}$ is composed of the energy-momentum tensor $T_{ab}$ of standard model
fields and of a $\lambda $ brane tension term:%
\begin{equation}
\tau _{ab}=-\lambda {^{(s+1)\!\!}}g_{ab}+T_{ab}\ .  \label{tau}
\end{equation}%
We allow for a completely generic brane, with energy density $\rho $, \
homogeneous pressure $p$, tensor of anisotropic pressures $\Pi _{ab}$ and
energy transport $Q_{a}$:%
\begin{equation}
T_{ab}=\rho n_{a}n_{b}+pg_{ab}+\Pi _{ab}+2n_{(a}Q_{b)}\ .  \label{T}
\end{equation}

We would like to write the Lanczos equation (\ref{Lanczos1}) in terms of the
new variables introduced in the previous sections. For this we employ the
decomposition (\ref{L4abdecomp}) of the extrinsic curvature of $\mathcal{M}$%
. We find the following projections of the Lanczos equation: 
\begin{eqnarray}
\Delta L_{ab} &=&-\widetilde{\kappa }^{2}\left[ pg_{ab}+\Pi _{ab}+\frac{1}{s}%
g_{ab}\left( \lambda -3p+\rho \right) \right] \ ,  \label{DL} \\
\Delta \mathcal{K}_{a} &=&-\widetilde{\kappa }^{2}Q_{a}\ ,  \label{DcalK} \\
s\Delta \mathcal{L} &=&-\widetilde{\kappa }^{2}\left[ -\lambda +3p+\left(
s-1\right) \rho \right] \ .  \label{DcalL}
\end{eqnarray}%
We see that among the gravitational degrees of freedom only $\mathcal{K}_{a}$
will have a jump across the brane due to the distributional energy-momentum
tensor. This occurs only when there is energy transport on the brane. 
\textit{Thus in all cosmological models with perfect fluid on the brane none
of the dynamical variables characterizing gravity will have a jump across
the brane.}

\section{Intrinsic curvatures}

The Riemann tensor, expression of the intrinsic curvature of the $\Sigma
_{t\chi }$ hypersurfaces arises from the non-commutativity of the covariant
derivative $D$: 
\begin{equation}
R^{a}{}_{bcd}v^{b}=\left( D_{c}D_{d}-D_{d}D_{c}\right) v^{a}\ .  \label{Rim}
\end{equation}%
Here $v^{a}\in T\Sigma _{t,\chi }$ is arbitrary. The Riemann tensor of $%
\mathcal{B}$ is defined in an analogous manner: 
\begin{equation}
\widetilde{R}^{a}{}_{bcd}\widetilde{v}_{b}=(\widetilde{\nabla }_{c}%
\widetilde{\nabla }_{d}-\widetilde{\nabla }_{d}\widetilde{\nabla }_{c})%
\widetilde{v}^{a}\ ,  \label{tildeRim}
\end{equation}%
for any $\widetilde{v}^{a}\in T\mathcal{B}.$ Straightforward computation
leads to the Gauss equation: 
\begin{eqnarray}
R^{a}{}_{bcd} &=&g_{i}^{a}{}g_{b}^{j}{}g_{c}^{k}{}g_{d}^{l}{}\widetilde{R}%
^{i}{}_{jkl}-K^{a}{}_{c}K_{bd}+K^{a}{}_{d}K_{bc}  \notag \\
&&+L^{a}{}_{c}L_{bd}-L^{a}{}_{d}L_{bc}\ .  \label{Gauss}
\end{eqnarray}%
This standard result holds for any $s$-dimensional hypersurface embedded in
an $(s+2)$-dimensional space \cite{Schouten}.

All other independent projections of $\widetilde{R}^{a}{}_{bcd}$ are
enlisted in Appendix C.

By contracting the Gauss equation twice, the relation between the scalar
curvatures emerges: 
\begin{eqnarray}
R{} &=&\widetilde{R}{}+2\widetilde{R}_{ab}{}(n^{a}n^{b}-l^{a}l^{b})-2%
\widetilde{R}_{abcd}{}n^{a}l^{b}n^{c}l^{d}  \notag \\
&&-K^{2}+K^{ab}{}K_{ab}+L^{2}-L^{ab}{}L_{ab}\ .  \label{Gauss2}
\end{eqnarray}%
Employing 
\begin{subequations}
\label{Ricciprojdiv}
\begin{eqnarray}
n^{a}n^{b}\widetilde{R}_{ab} &=&\widetilde{\nabla }_{a}(\alpha ^{a}-n^{a}%
\widetilde{\nabla }_{b}n^{b})+K^{2}-K_{ab}K^{ab}  \notag \\
&&-2\mathcal{K}_{a}\mathcal{K}^{a}+2K\mathcal{K}\ , \\
l^{a}l^{b}\widetilde{R}_{ab} &=&\widetilde{\nabla }_{a}(\lambda ^{a}-l^{a}%
\widetilde{\nabla }_{b}l^{b})+L^{2}-L_{ab}L^{ab}  \notag \\
&&+2\mathcal{K}_{a}\mathcal{K}^{a}-2L\mathcal{L}\ , \\
n^{a}{}l^{b}n^{c}l^{d}\widetilde{R}_{abcd} &=&\mathcal{L}^{2}-\mathcal{K}%
^{2}-3\mathcal{K}_{a}\mathcal{K}^{a}  \notag \\
&&-l^{a}\widetilde{\nabla }_{a}\mathcal{L}-n^{a}\widetilde{\nabla }_{a}%
\mathcal{K}-{}\alpha ^{b}\lambda _{b}\ .
\end{eqnarray}%
we find the relation between the $s$- and $(s+2)$-dimensional curvature
scalars: 
\end{subequations}
\begin{eqnarray}
R &=&\widetilde{R}+K^{2}-K_{ab}K^{ab}-2\mathcal{K}_{a}\mathcal{K}^{a}+2%
\mathcal{K}K  \notag \\
&&-L^{2}+L_{ab}L^{ab}+2\mathcal{L}L+2\alpha ^{b}\lambda _{b}  \notag \\
&&+2\widetilde{\nabla }_{a}(\alpha ^{a}-\lambda ^{a}-Kn^{a}+Ll^{a})\ .
\label{curvscalar}
\end{eqnarray}%
Appendix D contains an independent derivation of Eq. (\ref{curvscalar}).
Keeping in mind $\mathcal{K}_{a}=\mathcal{L}_{a}$, this equation shows a
perfect symmetry between quantities related to $l^{a}$ and $n^{a}$. The
expressions of the $\left( s+1+1\right) $-decomposed Riemann, Ricci and
Einstein tensors in terms of tensors on $\Sigma _{t\chi }$ are also given in
Appendix D, as well as the relation between the scalar curvatures in a form
without total divergences.

\section{Time-derivatives of extrinsic curvatures}

The energy-momentum tensor $\widetilde{T}_{ab}$ of the bulk receives two
types of contributions: $\widetilde{\Pi }_{ab}$ from the non-standard-model
fields in the bulk and $\tau _{ab}$ from standard-model fields on the brane.
The latter is a distributional source located on the brane, at $\chi =0$ (or
in terms of generic coordinates $\widetilde{x}^{a}$ given covariantly as $%
\chi \left( \left\{ \widetilde{x}^{a}\right\} \right) =0$).  In the special
case when the bulk contains only a cosmological constant $\widetilde{\Lambda 
}$, but with a generic source on the brane, from Eqs. (\ref{tau}) and (\ref%
{T}) we obtain%
\begin{eqnarray}
\widetilde{T}_{ab} &=&-\widetilde{\kappa }^{2}\widetilde{\Lambda }l_{a}l_{b}+%
\left[ \left( \rho +\lambda \right) \delta \left( \chi \right) +\widetilde{%
\kappa }^{2}\widetilde{\Lambda }\right] n_{a}n_{b}  \notag \\
&&+\left[ \left( p-\lambda \right) \delta \left( \chi \right) -\widetilde{%
\kappa }^{2}\widetilde{\Lambda }\right] g_{ab}  \notag \\
&&+\left[ \Pi _{ab}+2n_{(a}Q_{b)}\right] \delta \left( \chi \right) \ ,
\label{tildeT}
\end{eqnarray}%
and 
\begin{equation}
\widetilde{T}=-5\widetilde{\kappa }^{2}\widetilde{\Lambda }-\left( \rho
-3p+4\lambda \right) \delta \left( \chi \right) \ .
\end{equation}%
By employing the bulk Einstein equation%
\begin{equation}
\widetilde{R}_{ab}=\widetilde{T}_{ab}-\frac{\widetilde{T}}{s}\widetilde{g}%
_{ab}\ ,
\end{equation}%
we obtain the following projections of the bulk Ricci tensor: 
\begin{subequations}
\begin{eqnarray}
&&g_{a}^{c}{}g_{b}^{d}\widetilde{R}_{cd}=\frac{5-s}{s}\widetilde{\kappa }^{2}%
\widetilde{\Lambda }g_{ab}  \notag \\
&&+\left[ \frac{\rho +\left( s-3\right) p+\left( 4-s\right) \lambda }{s}%
g_{ab}+\Pi _{ab}\right] \delta \left( \chi \right) \ , \\
&&g_{a}^{c}{}l^{d}\widetilde{R}_{cd}=0\ , \\
&&l^{a}l^{b}\widetilde{R}_{ab}=\frac{5-s}{s}\widetilde{\kappa }^{2}%
\widetilde{\Lambda }+\frac{\rho -3p+4\lambda }{s}\delta \left( \chi \right)
\ .
\end{eqnarray}%
Employing these in Eqs. (\ref{ggRicci}), (\ref{Codazzi2}) and (\ref{llRicci}%
) of Appendix D and transforming the Lie-derivatives $\mathcal{L}_{\mathbf{n}%
}K_{ab},\ \mathcal{L}_{\mathbf{n}}\mathcal{K}_{a}$ and $\mathcal{L}_{\mathbf{%
n}}\mathcal{K}$ to time derivatives by Eq. (\ref{derivt}), the time
evolution of $K_{ab}$, $\mathcal{K}_{a}$ and $\mathcal{K}$ can be readily
deduced: 
\end{subequations}
\begin{subequations}
\label{extrcurvdot}
\begin{eqnarray}
\frac{\partial }{\partial t}K_{ab} &=&N\Biggl[g_{a}^{c}{}g_{b}^{d}{}%
\widetilde{R}_{cd}-R_{ab}+L_{ab}\left( L-\mathcal{L}\right) -2L_{ac}L_{b}^{c}
\notag \\
&&-K_{ab}\left( K+\mathcal{K}\right) +2K_{ac}K_{b}^{c}+2\mathcal{K}_{a}%
\mathcal{K}_{b}+\mathcal{L}_{\mathbf{l}}L_{ab}  \notag \\
&&+\frac{D_{b}D_{a}M}{M}\Biggr]+D_{b}D_{a}N+\mathcal{L}_{\mathbf{N}}K_{ab}\ ,
\\
\frac{\partial }{\partial t}\mathcal{K}_{a} &=&N\left[ -D^{b}L_{ab}+D_{a}%
\left( L-\mathcal{L}\right) -K\mathcal{K}_{a}\right] +\mathcal{L}_{\mathbf{N}%
}\mathcal{K}_{a}  \notag \\
&&-\left( L_{a}^{b}+\mathcal{L}\delta _{a}^{b}\right) D_{b}N\ , \\
\frac{\partial }{\partial t}\mathcal{K} &=&N\Biggl[l^{a}l^{b}\widetilde{R}%
_{ab}-L_{ab}L^{ab}+\mathcal{L}^{2}+\frac{D_{a}D^{a}M}{M}  \notag \\
&&-2\mathcal{K}_{a}\mathcal{K}^{a}-\mathcal{K}\left( K+\mathcal{K}\right) +%
\mathcal{L}_{\mathbf{l}}\left( L-\mathcal{L}\right) \Biggr]  \notag \\
&&+\frac{D^{a}M}{M}D_{a}N+\mathcal{L}_{\mathbf{N}}\mathcal{K}\ .
\end{eqnarray}%
The corresponding expressions containing $\chi $-derivatives (which are
similar to those of the time-evolution equations (\ref{Kab}), (\ref{Kcal})
and (\ref{Kacal}) of the complementary set of dynamical data) are: 
\end{subequations}
\begin{subequations}
\label{extrcurvdotdet}
\begin{eqnarray}
\frac{\partial }{\partial t}K_{ab} &=&N\Biggl[g_{a}^{c}{}g_{b}^{d}{}%
\widetilde{R}_{cd}-R_{ab}+L_{ab}\left( L-\mathcal{L}\right) -2L_{ac}L_{b}^{c}
\notag \\
&&-K_{ab}\left( K+\mathcal{K}\right) +2K_{ac}K_{b}^{c}+2\mathcal{K}_{a}%
\mathcal{K}_{b}+\frac{D_{b}D_{a}M}{M}  \notag \\
&&+\frac{1}{M}\left( \frac{\partial }{\partial \chi }%
L_{ab}-M^{c}D_{c}L_{ab}-2L_{c(a}D_{b)}M^{c}\right) \Biggr]  \notag \\
&&+D_{b}D_{a}N+N^{c}D_{c}K_{ab}+2K_{c(a}D_{b)}N^{c}\ , \\
\frac{\partial }{\partial t}\mathcal{K}_{a} &=&N\left[ -D^{b}L_{ab}+D_{a}%
\left( L-\mathcal{L}\right) -K\mathcal{K}_{a}\right] +N^{b}D_{b}\mathcal{K}%
^{a}  \notag \\
&&-\left( L_{a}^{b}+\mathcal{L}\delta _{a}^{b}\right) D_{b}N+\mathcal{K}%
_{b}D_{a}N^{b}\ , \\
\frac{\partial }{\partial t}\mathcal{K} &=&N\Biggl\{l^{a}l^{b}\widetilde{R}%
_{ab}-L_{ab}L^{ab}+\mathcal{L}^{2}+\frac{D_{a}D^{a}M}{M}  \notag \\
&&-2\mathcal{K}_{a}\mathcal{K}^{a}-\mathcal{K}\left( K+\mathcal{K}\right)  
\notag \\
&&+\frac{1}{M}\left[ \frac{\partial }{\partial \chi }\left( L-\mathcal{L}%
\right) -M^{a}D_{a}\left( L-\mathcal{L}\right) \right] \Biggr\}  \notag \\
&&+\frac{D^{a}M}{M}D_{a}N+N^{a}D_{a}\mathcal{K}\ .
\end{eqnarray}%
We have employed Eq. (\ref{derivchi}) in their derivation. Note that in the
above formulae $L_{ab}$ and $\mathcal{L}$ are also given in terms of
dynamical data, cf. Eqs. (\ref{Lab}) and (\ref{Lcal}).

\section{Concluding remarks}

We have developed a decomposition scheme of the $\left( s+2\right) $
dimensional space-time based of two perpendicular foliations of constant time
and constant $\chi $-surfaces. In a brane-world scenario the latter contains
the brane at $\chi =0$. A careful geometrical interpretation has allowed for
identifying dynamical quantities with geometrical expressions. From among
the various projections to $\Sigma _{t\chi }$ of the extrinsic curvature
pertinent to$\ $the off-brane normal solely $\mathcal{L}^{a}$ was found to
be dynamical, together with \textit{all }components ($K_{ab}$, $\mathcal{K}%
^{a}=\mathcal{L}^{a}$ and $\mathcal{K}$) of the extrinsic curvature related
to $n^{a}$. Their expression was given in terms of time derivatives of the
metric $g_{ab\text{ }}$induced on $\Sigma _{t\chi }$, shift vector
components $M^{a}$ and lapse $M$ of $\partial /\partial \chi $ . Time
evolution of the second fundamental form $K_{ab}$, of the normal fundamental
form $\mathcal{K}^{a}$ and normal fundamental scalar $\mathcal{K}$ were also
derived for a generic brane. The Lanczos equation was writen in terms of the
same set of variables.

Our formalism applied for $s=2$ is different from previous $2+1+1$
decompositions in general relativity, developed to deal with stationary and
axisymmetric space-times. There the temporal and spatial directions singled
out are a stationary and a rotational Killing vector. By contrast, we are
interested in \textit{evolutions} along the singled-out timelike and
off-brane directions. The formalism developed in \cite{Maeda1}, \cite{Maeda2}
relies on the use of a factor space with respect to the rotational Killing
vector. The induced metric is then defined with this Killing vector and the
formalism becomes rather complicated. In a more recent approach \cite{GB},
Gourgoulhon and Bonazzola introduce the induced metrics by using normal
vector fields to the $2$-space (like we do), hence they avoid the use of
twist-related quantities. However their treatment relies on first
decomposing space-time with respect to the temporal direction, next with
respect to the spatial direction, and this procedure unfortunately lets no
counterpart to the brane extrinsic curvature ${^{(s+1)\!\!}}L_{ab}$ in their
formalism. Thus their formalism is not suited for a generalization to
brane-world scenarios, where the Lanczos equation is given precisely in
terms of ${^{(s+1)\!\!}}L_{ab}$. For convenience we give a comparative table
of the quantities appearing in the approaches of \cite{GB} and ours (Table %
\ref{Table1}).

\begin{widetext}
\begin{table}[h]
\caption{Comparison of our notations and quantities employed with those of
Ref. \protect\cite{GB}. A triple dot denotes the absence of the respective
quantity from the formalism. We note that the correspondences $%
l^{a}\leftrightarrow m^{\protect\alpha }$ and $L_{ab}\leftrightarrow L_{%
\protect\alpha \protect\beta }$ hold only because we have chosen the two
foliations perpendicular to each other.}
\begin{equation*}  \label{Table1}
\begin{tabular}{l|l|l}
& our formalism & Gourgoulhon and Bonazzola \\ \hline\hline
manifolds & $\left( \mathcal{B}\text{, }\mathcal{M}_{\chi }\text{, }S_{t}%
\text{, }\Sigma _{t\chi }\right) $ & $\left( \mathcal{E}\text{, }...\text{, }%
\Sigma _{t}\text{, }\Sigma _{t\phi }\text{ }\right) $ \\ 
metrics & $\left( \widetilde{g}\text{, }{^{(s+1)\!}g}\text{, }\widehat{g}%
\text{,}\ g\right) $ & $\left( g\text{, }...\text{, }h\text{, }k\right) $ \\ 
metric compatible covariant derivatives & $\left( \widetilde{\nabla }\text{, 
}^{(s+1)\!\!\!}D\text{,\ }\widehat{D}\text{, }D\right) $ & $\left( _{;}\text{%
, }...\text{, }_{\mid }\text{, }_{\parallel }\right) $ \\ 
coordinates & $\left( t\text{, }\chi \right) $ & $\left( t\text{, }\varphi
\right) $ \\ 
singled-out vectors & $\left( \frac{\partial }{\partial t}\text{, }\frac{%
\partial }{\partial \chi }\right) $ & $\left( \frac{\partial }{\partial t}%
\text{, }\frac{\partial }{\partial \varphi }\right) $ \\ 
normals & $\left( n^{a}\text{, }l^{a}\right) $ & $\left( n^{\alpha }\text{, }%
m^{\alpha }\right) $ \\ 
accelerations & $\left( \alpha ^{a}\text{, }\widehat{g}_{b}^{a}\lambda
^{b}\right) $ & $\left( a^{\alpha }\text{, }b^{\alpha }\right) $ \\ 
shifts & $\left( N^{a}\text{, }\mathcal{N}\right) $, $M^{a}$ & $-N^{\alpha
}=-\left( q^{\alpha },\omega \right) $, $-M^{\alpha }$ \\ 
extrinsic curvatures & $\left( ^{(s+1)\!\!\!}L_{ab}\text{,\ }\widehat{K}%
_{ab}\right) $ & $\left( ...\text{,\ }-K_{\alpha \beta }\right) $ \\ 
extrinsic curvature projections & $\left( K_{ab}\text{, }L_{ab}\text{, }%
\mathcal{K}_{a}=\mathcal{L}_{a}\text{, }\mathcal{K}\text{, }\mathcal{L}%
\right) $ & $\left( ...\text{, }-L_{\alpha \beta }\text{, }...\text{, }...%
\text{, ...}\right) $%
\end{tabular}%
\end{equation*}%
\end{table}
\end{widetext}

The system of equations giving the evolution of $\left( g_{ab}\text{, }M^{a}%
\text{, }M\right) $ and $\left( K_{ab}\text{, }\mathcal{K}^{a}\text{, }%
\mathcal{K}\right) $ represents the gravitational dynamics in terms of
variables adapted to the brane. In top of these there are constraints on
their initial values to be satisfied. These are the Hamiltonian and
diffeomorphism constraints. Their derivation from a variational principle in
terms of our brane-adapted variables, keeping in mind that some of the
dynamics is frozen due to the existence of the brane as a hypersurface, is
in progress and will be published in a forthcoming paper.
\end{subequations}

\section{Acknowledgments}

This work was supported by OTKA grants no. T046939 and TS044665. L.\'A.G.
was further supported by the J\'{a}nos Bolyai Scholarship of the Hungarian
Academy of Sciences.

\appendix

\section{The two foliations}

The $t$=constant hypersurfaces $S_{t}$ are defined by the one-form field $%
\overline{n}=Ndt$. Similarly, $\mathcal{M}_{\chi }$ is the hypersurfaces $%
\chi $=constant, defined by $\overline{l}=M^{\prime }d\chi $. We introduce
the one-form field $\overline{m}$ such that the metric in $\mathcal{B}$ can
be written as: 
\begin{equation}
\widetilde{g}=g_{\mathbf{ij}}F^{\mathbf{i}}\otimes F^{\mathbf{j}}+\overline{m%
}\otimes \overline{m}-\overline{n}\otimes \overline{n}\ .  \label{tildeg}
\end{equation}%
Here the co-basis $\left\{ f^{\mathbf{\alpha }}\right\} =\{\overline{n},F^{%
\mathbf{i}},\overline{m}\}$ has the dual basis $\left\{ f_{\mathbf{\beta }%
}\right\} =\{n,F_{\mathbf{j}},m\}$, where $\{F_{\mathbf{j}}\}$ is some basis
in $T\Sigma _{t\chi }$. Thus $n_{a}=-\overline{n}_{a}$ and $m_{a}=\overline{m%
}_{a}$. The normal vector $l^{a}$ to $\mathcal{M}_{\chi }$ can be
conveniently parameterized as%
\begin{equation}
l^{a}=n^{a}\sinh \gamma +m^{a}\cosh \gamma \ .
\end{equation}%
This parameterization assures that $l^{a}$ has unit norm with respect to the
metric (\ref{tildeg}). The dual form to $l^{a}$ is%
\begin{equation}
\overline{l}=\widetilde{g}\left( .,l\right) =-\overline{n}\sinh \gamma +%
\overline{m}\cosh \gamma \ .
\end{equation}%
We introduce the second basis $\left\{ e_{\mathbf{\alpha }}\right\}
=\{\partial /\partial t,E_{\mathbf{i}}=F_{\mathbf{i}},\partial /\partial
\chi \}$ and its dual co-basis $\left\{ e^{\mathbf{\beta }}\right\} =\{dt,E^{%
\mathbf{i}},d\chi \}$. The two sets of bases are related as: 
\begin{eqnarray}
\overline{n} &=&Ndt\ ,\quad F^{\mathbf{i}}=A_{\mathbf{\beta }}^{\mathbf{i}%
}e^{\mathbf{\beta }}\ ,  \notag \\
\overline{m} &=&\frac{M^{\prime }d\chi +N\sinh \gamma dt}{\cosh \gamma }\ ,
\end{eqnarray}%
and 
\begin{equation}
\frac{\partial }{\partial t}=N^{\mathbf{\beta }}f_{\mathbf{\beta }}\ ,\quad 
\frac{\partial }{\partial \chi }=M^{\mathbf{\beta }}f_{\mathbf{\beta }}\
,\quad E_{\mathbf{i}}=F_{\mathbf{i}}\ .
\end{equation}%
The unknown coefficients $A_{\mathbf{\beta }}^{\mathbf{i}}$, $N^{\mathbf{%
\beta }}$ and $M^{\mathbf{\beta }}$ are constrained by the duality relations 
$\left\langle e^{\mathbf{\alpha }},e_{\mathbf{\beta }}\right\rangle =\delta
_{\mathbf{\beta }}^{\mathbf{\alpha }}=\left\langle f^{\mathbf{\alpha }},f_{%
\mathbf{\beta }}\right\rangle $. After some algebra we find the relation
between the two co-bases 
\begin{eqnarray}
\overline{n} &=&Ndt\ ,  \notag \\
\overline{m} &=&\mathcal{N}dt+Md\chi \ ,  \notag \\
F^{\mathbf{i}} &=&N^{\mathbf{i}}dt+E^{\mathbf{i}}+M^{\mathbf{i}}d\chi \ ,
\label{fdualbasis}
\end{eqnarray}%
and the relation between the bases 
\begin{subequations}
\begin{eqnarray}
\frac{\partial }{\partial t} &=&Nn+N^{\mathbf{i}}F_{\mathbf{i}}+\mathcal{N}%
m\ ,  \label{ddt} \\
\frac{\partial }{\partial \chi } &=&M^{\mathbf{i}}F_{\mathbf{i}}+Mm\ ,
\label{ddchi} \\
E_{\mathbf{i}} &=&F_{\mathbf{i}}\ .
\end{eqnarray}%
We have introduced the shorthand notations 
\end{subequations}
\begin{equation}
\mathcal{N}=N\tanh \gamma \ ,\quad M=\frac{M^{\prime }}{\cosh \gamma }\ .
\end{equation}%
Eq. (\ref{ddt}) shows that the the time-evolution vector $\partial /\partial
t$ is decomposed as in the ADM\ formalism of general relativity: $N$ is the
lapse function and $(N^{\mathbf{i}},\mathcal{N})$ are the 4D shift vector
components in the chosen basis. The functions $(M^{\mathbf{i}},M)\,$\
represent the arbitrary tangential and normal contributions to the off-brane
evolution vector $\partial /\partial \chi $ with respect to $\Sigma _{t\chi
} $. Eq. (\ref{ddchi}) shows that there is no $n$-term in $\partial
/\partial \chi $, thus off-brane evolution in the coordinate $\chi $ happens
in $S_{t}$, thus at constant time.

For convenience we also give the inverted relations among the bases: 
\begin{eqnarray}
dt &=&\frac{\overline{n}}{N}\ ,  \notag \\
d\chi &=&\frac{1}{M}\left( -\frac{\mathcal{N}}{N}\overline{n}+\overline{m}%
\right) \ ,  \notag \\
E^{\mathbf{j}} &=&\frac{1}{N}\left( \frac{\mathcal{N}}{M}M^{\mathbf{j}}-N^{%
\mathbf{j}}\right) \overline{n}+F^{\mathbf{j}}-\frac{M^{\mathbf{j}}}{M}%
\overline{m}\ ,  \label{edualbasis}
\end{eqnarray}%
and 
\begin{eqnarray}
n &=&\frac{1}{N}\left[ \frac{\partial }{\partial t}+\left( \frac{\mathcal{N}%
}{M}M^{\mathbf{i}}-N^{\mathbf{i}}\right) E_{\mathbf{i}}-\frac{\mathcal{N}}{M}%
\frac{\partial }{\partial \chi }\right] \ .  \notag \\
m &=&\frac{1}{M}\left( -M^{\mathbf{i}}E_{\mathbf{i}}+\frac{\partial }{%
\partial \chi }\right) \ ,  \notag \\
F_{\mathbf{i}} &=&E_{\mathbf{i}}\ .  \label{fbasis}
\end{eqnarray}

Obviously the two foliations will become perpendicular for $\gamma =0=%
\mathcal{N}$. Then the vectors $l$ and $m$ coincide.

\begin{widetext}

\section{Decomposition of the metric}

By substituting Eq. (\ref{fdualbasis}) into the $\left( s+2\right) $-metric (%
\ref{tildeg}) we obtain 
\begin{equation}
\widetilde{g}_{\mathbf{\alpha \beta }}=\left( 
\begin{array}{ccc}
g_{\mathbf{ij}}N^{\mathbf{i}}N^{\mathbf{j}}+\mathcal{N}^{2}-N^{2} & g_{%
\mathbf{ij}}N^{\mathbf{j}} & g_{\mathbf{ij}}N^{\mathbf{i}}M^{\mathbf{j}}+%
\mathcal{N}M \\ 
g_{\mathbf{ij}}N^{\mathbf{i}} & g_{\mathbf{ij}} & g_{\mathbf{ij}}M^{\mathbf{i%
}} \\ 
g_{\mathbf{ij}}N^{\mathbf{i}}M^{\mathbf{j}}+\mathcal{N}M & g_{\mathbf{ij}}M^{%
\mathbf{j}} & g_{\mathbf{ij}}M^{\mathbf{i}}M^{\mathbf{j}}+M^{2}%
\end{array}%
\right) \ .  \label{tildeg1}
\end{equation}%
To proove in a simple way the relationship (\ref{gdet}) between the
determinants of the $(s+2)$- and $s$-metrics, we transform the determinant $%
\widetilde{g}$ by suitably combining the columns and lines as follows: 
\begin{eqnarray}
\widetilde{g} &=&\left| 
\begin{array}{ccc}
\mathcal{N}^{2}-N^{2} & g_{\mathbf{ij}}N^{\mathbf{j}} & g_{\mathbf{ij}}N^{%
\mathbf{i}}M^{\mathbf{j}}+\mathcal{N}M \\ 
0 & g_{\mathbf{ij}} & g_{\mathbf{ij}}M^{\mathbf{i}} \\ 
\mathcal{N}M & g_{\mathbf{ij}}M^{\mathbf{j}} & g_{\mathbf{ij}}M^{\mathbf{i}%
}M^{\mathbf{j}}+M^{2}%
\end{array}%
\right|  \notag \\
&=&[\mathcal{N}^{2}-N^{2}]\left| 
\begin{array}{cc}
g_{\mathbf{ij}} & g_{\mathbf{ij}}M^{\mathbf{i}} \\ 
g_{\mathbf{ij}}M^{\mathbf{j}} & g_{\mathbf{ij}}M^{\mathbf{i}}M^{\mathbf{j}%
}+M^{2}%
\end{array}%
\right| +\mathcal{N}M\left| 
\begin{array}{cc}
g_{\mathbf{ij}}N^{\mathbf{j}} & g_{\mathbf{ij}}N^{\mathbf{i}}M^{\mathbf{j}}+%
\mathcal{N}M \\ 
g_{\mathbf{ij}} & g_{\mathbf{ij}}M^{\mathbf{i}}%
\end{array}%
\right|  \notag \\
&=&[\mathcal{N}^{2}-N^{2}]\left| 
\begin{array}{cc}
g_{\mathbf{ij}} & g_{\mathbf{ij}}M^{\mathbf{i}} \\ 
0 & M^{2}%
\end{array}%
\right| +\mathcal{N}M\left| 
\begin{array}{cc}
0 & \mathcal{N}M \\ 
g_{\mathbf{ij}} & g_{\mathbf{ij}}M^{\mathbf{i}}%
\end{array}%
\right| =-N^{2}M^{2}g\ .
\end{eqnarray}%
The inverse of the metric (\ref{tildeg1}) has a more cumbersome expression: 
\begin{equation}
\widetilde{g}^{\alpha \beta }=\frac{1}{N^{2}}\left( 
\begin{matrix}
\displaystyle{-1} & \displaystyle{N^{\mathbf{i}}-\frac{\mathcal{N}}{M}M^{%
\mathbf{i}}} & \displaystyle{\frac{\mathcal{N}}{M}} \\ 
&  &  \\ 
\displaystyle{N^{\mathbf{j}}-\frac{\mathcal{N}}{M}M^{\mathbf{j}}} & %
\displaystyle{N^{2}g^{\mathbf{ij}}-N^{\mathbf{i}}N^{\mathbf{j}}+\frac{N^{2}-%
\mathcal{N}^{2}}{M^{2}}M^{\mathbf{i}}M^{\mathbf{j}}+2\frac{\mathcal{N}}{M}%
N^{(\mathbf{i}}M^{\mathbf{j})}} & \displaystyle{\frac{\mathcal{N}^{2}-N^{2}}{%
M^{2}}M^{\mathbf{j}}-\frac{\mathcal{N}}{M}N^{\mathbf{j}}} \\ 
&  &  \\ 
\displaystyle{\frac{\mathcal{N}}{M}} & \displaystyle{\frac{\mathcal{N}%
^{2}-N^{2}}{M^{2}}M^{\mathbf{i}}-\frac{\mathcal{N}}{M}N^{\mathbf{i}}} & %
\displaystyle{\frac{N^{2}-\mathcal{N}^{2}}{M^{2}}}%
\end{matrix}%
\right) \ .  \label{tildeginv}
\end{equation}

It is convenient to free ourselves from the particular bases employed above.
For this purpose we define the generic tensorial expressions $g_{ab}=g_{%
\mathbf{ij}}\left( F^{\mathbf{i}}\right) _{a}\left( F^{\mathbf{j}}\right)
_{b}\ ,\ N^{a}=N^{\mathbf{i}}\left( F_{\mathbf{i}}\right) ^{a}\ ,\ M^{a}=M^{%
\mathbf{i}}\left( F_{\mathbf{i}}\right) ^{a}$, in terms of which Eqs. (\ref%
{tildeg}), (\ref{ddt}) and (\ref{ddchi}) give Eqs. (\ref{tildeg0}, (\ref%
{ddt0}) and (\ref{ddchi0}).

\section{Curvatures of normal congruences}

We start with the computation of the nontrivial Lie-brackets of the basis
vectors $\{f_{\mathbf{\beta }}\}$ $=\{n,F_{\mathbf{j}}=\partial /\partial y^{%
\mathbf{j}},m\}$, employing first the relations (\ref{fbasis}), second that $%
\{e_{\mathbf{\beta }}\}$ is a coordinate basis and finally rewriting the
resulting expressions in the $\{f_{\mathbf{\beta }}\}$ basis: 
\begin{subequations}
\begin{eqnarray}
\lbrack n,F_{\mathbf{j}}]^{a} &=&\partial _{\mathbf{j}}\left( \ln N\right)
n^{a}+\frac{1}{N}\left( \partial _{\mathbf{j}}N^{\mathbf{i}}-\frac{\mathcal{N%
}}{M}\partial _{\mathbf{j}}M^{\mathbf{i}}\right) \left( F_{\mathbf{i}%
}\right) ^{a}+\frac{M}{N}\partial _{\mathbf{j}}\left( \frac{\mathcal{N}}{M}%
\right) m^{a}\ ,  \label{nF} \\
\lbrack m,F_{\mathbf{j}}]^{a} &=&\partial _{\mathbf{j}}\left( \ln M\right)
m^{a}+\frac{1}{M}\partial _{\mathbf{j}}M^{\mathbf{i}}\left( F_{\mathbf{i}%
}\right) ^{a}\ ,  \label{lF} \\
\lbrack n,m]^{a} &=&\frac{1}{M}\left[ \frac{\partial }{\partial \chi }\left(
\ln N\right) -M^{\mathbf{j}}\partial _{\mathbf{j}}\left( \ln N\right) \right]
n^{a}+\frac{1}{MN}\left[ -\frac{\partial }{\partial t}M+\frac{\partial }{%
\partial \chi }\mathcal{N}-M^{\mathbf{j}}\partial _{\mathbf{j}}\mathcal{N}%
+N^{\mathbf{j}}\partial _{\mathbf{j}}M\right] m^{a}  \notag \\
&&+\frac{1}{MN}\left[ -\widetilde{\mathcal{L}}_{\frac{\partial }{\partial t}%
}M^{\mathbf{i}}+\widetilde{\mathcal{L}}_{\frac{\partial }{\partial \chi }}N^{%
\mathbf{i}}-M^{\mathbf{j}}\partial _{\mathbf{j}}N^{\mathbf{i}}+N^{\mathbf{j}%
}\partial _{\mathbf{j}}M^{\mathbf{i}}\right] \left( F_{\mathbf{i}}\right)
^{a}\ .  \label{nl}
\end{eqnarray}%
However Frobenius theorem states that $[n,F_{\mathbf{j}}]^{a}$ should have
no $m^{a}$ component. Therefore we get the condition 
\end{subequations}
\begin{equation}
\mathcal{N}=\nu M\ ,\quad D_{a}\nu =0\ .  \label{Frobenius}
\end{equation}%
Thus $\mathcal{N}$ should be proportional to $M$, with a proportionality
coefficient, which is constant along $\Sigma _{t\chi }$.

Next we announce the following:

\textbf{Theorem. }If $\widetilde{\nabla }$ is the connection compatible with
the metric $\widetilde{g}$, any set of vectors $\{f_{\mathbf{\alpha }}\}$
obeying $\widetilde{g}\left( f_{\mathbf{\alpha }},f_{\mathbf{\beta }}\right) 
$=const. also satisfy the relation $\widetilde{g}\left( f_{\mathbf{\alpha }},%
\widetilde{\nabla }_{\mathbf{f}_{\mathbf{\beta }}}f_{\mathbf{\beta }}\right)
=$ $\widetilde{g}\left( \left[ f_{\mathbf{\alpha }},f_{\mathbf{\beta }}%
\right] ,f_{\mathbf{\beta }}\right) $.

\textit{Proof. }$\widetilde{g}\left( f_{\mathbf{\alpha }},\widetilde{\nabla }%
_{\mathbf{f}_{\mathbf{\beta }}}f_{\mathbf{\beta }}\right) =-\widetilde{g}%
\left( \widetilde{\nabla }_{\mathbf{f}_{\mathbf{\beta }}}f_{\mathbf{\alpha }%
},f_{\mathbf{\beta }}\right) =-\widetilde{g}\left( \widetilde{\nabla }_{%
\mathbf{f}_{\mathbf{\alpha }}}f_{\mathbf{\beta }},f_{\mathbf{\beta }}\right)
-\widetilde{g}\left( \left[ f_{\mathbf{\beta }},f_{\mathbf{\alpha }}\right]
,f_{\mathbf{\beta }}\right) =$ $\widetilde{g}\left( \left[ f_{\mathbf{\alpha 
}},f_{\mathbf{\beta }}\right] ,f_{\mathbf{\beta }}\right) $ Q.E.D.

Finally we employ the above result for $f_{\mathbf{\beta }}$ either $n$ or $%
m $ together with the commutators (\ref{nF})-(\ref{nl}) to obtain explicit
expressions for various components of the curvatures $\alpha ^{a}$ and $%
\lambda ^{a}$: 
\begin{subequations}
\begin{eqnarray}
\alpha _{b} &=&\widetilde{g}\left( f_{\mathbf{\alpha }},\widetilde{\nabla }_{%
\mathbf{n}}n\right) \left( f^{\mathbf{\alpha }}\right) _{b}=D_{b}\left( \ln
N\right) +\frac{1}{M}\left[ \frac{\partial }{\partial \chi }\left( \ln
N\right) -M^{a}D_{a}\left( \ln N\right) \right] m_{b}\ ,  \label{accdet} \\
\lambda _{b} &=&\widetilde{g}\left( f_{\mathbf{\alpha }},\widetilde{\nabla }%
_{\mathbf{l}}l\right) \left( f^{\mathbf{\alpha }}\right) _{b}=-D_{b}\left(
\ln M\right) +\frac{1}{MN}\left[ \frac{\partial }{\partial t}M-\frac{%
\partial }{\partial \chi }\mathcal{N}+M^{a}D_{a}\mathcal{N}-N^{a}D_{a}M%
\right] n_{b}\ .  \label{lambdadet}
\end{eqnarray}%
\end{subequations}
We have employed the relations $D_{b}N=\left( F^{\mathbf{i}}\right)
_{b}\partial _{\mathbf{i}}N$ and $N^{a}\partial _{a}M=N^{\mathbf{i}}\partial
_{\mathbf{i}}M$, deducible from the coordinate basis character of $\left\{
e_{\beta }\right\} $.

\section{Decomposition of curvatures}

In this Appendix we enlist the relations among the $\left( s+2\right) $ and $%
s$-dimensional Riemann, Ricci and Einstein tensors, as well as the
relation between the scalar curvatures in a fully decomposed form. The
formulae are valid for perpendicular foliations, $m^{a}=l^{a}$.

The projections of the Riemann tensor are Eq. (\ref{Gauss}) and 
\begin{subequations}
\begin{eqnarray}
n^{i}g_{b}^{j}{}g_{c}^{k}{}g_{d}^{l}{}\widetilde{R}_{ijkl}
&=&D_{d}K_{bc}-D_{c}K_{bd}+\mathcal{K}{}_{c}L_{bd}-\mathcal{K}{}_{d}L_{bc}\ ,
\\
n^{i}{}g_{b}^{j}{}n^{k}g_{d}^{l}{}\widetilde{R}_{ijkl} &=&-\mathcal{L}_{%
\mathbf{n}}K_{bd}+K_{bk}K_{d}^{k}+\mathcal{K}_{b}\mathcal{K}_{d}-\mathcal{L}%
L_{bd}+N^{-1}D_{d}D_{b}N\ , \\
l^{i}g_{b}^{j}{}g_{c}^{k}{}g_{d}^{l}{}\widetilde{R}_{ijkl}
&=&D_{d}L_{bc}-D_{c}L_{bd}+\mathcal{K}{}_{c}K_{bd}-\mathcal{K}{}_{d}K_{bc}\ ,
\\
l^{i}{}g_{b}^{j}{}l^{k}g_{d}^{l}{}\widetilde{R}_{ijkl} &=&-\mathcal{L}_{%
\mathbf{l}}L_{bd}+L_{bk}L_{d}^{k}-\mathcal{K}_{b}\mathcal{K}_{d}+\mathcal{K}%
K_{bd}-M^{-1}D_{d}D_{b}M\ , \\
n^{i}l^{j}g_{c}^{k}{}g_{d}^{l}{}\widetilde{R}_{ijkl} &=&D_{d}{}\mathcal{K}%
_{c}-D_{c}{}\mathcal{K}_{d}+L_{ci}K_{d}^{i}-L_{di}K_{c}^{i}\ , \\
n^{i}{}g_{b}^{j}{}g_{c}^{k}{}l^{l}\widetilde{R}_{ijkl} &=&-D_{c}{}\mathcal{K}%
_{b}+\mathcal{L}_{\mathbf{l}}K_{bc}-K_{kd}L_{b}^{k}-\mathcal{K}L_{bc}-2M^{-1}%
\mathcal{K}_{(b}D_{c)}M\ , \\
n^{i}{}l^{j}g_{c}^{k}{}l^{l}\widetilde{R}_{ijkl} &=&\mathcal{L}_{\mathbf{l}}%
\mathcal{K}_{c}-D_{c}\mathcal{K}+\mathcal{K}%
_{k}L_{c}^{k}+M^{-1}K_{c}^{i}D_{i}M-M^{-1}\mathcal{K}D_{c}M\ , \\
n^{i}{}l^{j}n^{k}g_{d}^{l}{}\widetilde{R}_{ijkl} &=&-D_{d}\mathcal{L}-%
\mathcal{L}_{\mathbf{n}}\mathcal{K}_{d}-\mathcal{K}%
_{i}K_{d}^{i}-N^{-1}L_{d}^{i}D_{i}N-N^{-1}\mathcal{L}D_{d}N\ , \\
n^{i}{}l^{j}n^{k}l^{l}\widetilde{R}_{ijkl} &=&\mathcal{L}^{2}-\mathcal{K}%
^{2}-3\mathcal{K}_{i}\mathcal{K}^{i}-\mathcal{L}_{\mathbf{l}}\mathcal{L}-%
\mathcal{L}_{\mathbf{n}}\mathcal{K}+\left( NM\right) ^{-1}D^{i}ND_{i}M\ .
\end{eqnarray}%
Contractions and multiplications with the normal vectors of the above
formulae give, respectively 
\end{subequations}
\begin{subequations}
\begin{eqnarray}
g_{a}^{c}{}g_{b}^{d}{}\widetilde{R}_{cd} &=&R_{ab}+K_{ab}\left( K+\mathcal{K}%
\right) -2K_{ac}K_{b}^{c}+\mathcal{L}_{\mathbf{n}}K_{ab}-N^{-1}D_{b}D_{a}N 
\notag \\
&&-2\mathcal{K}_{a}\mathcal{K}_{b}-L_{ab}\left( L-\mathcal{L}\right)
+2L_{ac}L_{b}^{c}-\mathcal{L}_{\mathbf{l}}L_{ab}-M^{-1}D_{b}D_{a}M\ ,
\label{ggRicci} \\
n^{a}n^{b}\widetilde{R}_{ab} &=&-\mathcal{L}_{\mathbf{n}}\left( K+\mathcal{K}%
\right) +K_{ab}K^{ab}+N^{-1}D_{a}D^{a}N-2\mathcal{K}_{a}\mathcal{K}^{a}-%
\mathcal{K}^{2}  \notag \\
&&-\mathcal{L}_{\mathbf{l}}\mathcal{L}-\mathcal{L}\left( L-\mathcal{L}%
\right) +\left( NM\right) ^{-1}D_{a}ND^{a}M\ ,  \label{nnRicci} \\
l^{a}l^{b}\widetilde{R}_{ab} &=&-\mathcal{L}_{\mathbf{l}}\left( L-\mathcal{L}%
\right) +L_{ab}L^{ab}-M^{-1}D_{a}D^{a}M+2\mathcal{K}_{a}\mathcal{K}^{a}-%
\mathcal{L}^{2}  \notag \\
&&+\mathcal{L}_{\mathbf{n}}\mathcal{K}+\mathcal{K}\left( K+\mathcal{K}%
\right) -\left( NM\right) ^{-1}D_{a}ND^{a}M\ ,  \label{llRicci} \\
n^{a}l^{b}\widetilde{R}_{ab} &=&D_{a}\mathcal{K}^{a}-\mathcal{L}_{\mathbf{l}%
}K+K_{ab}L^{ab}+\mathcal{K}L+M^{-1}\mathcal{K}^{a}D_{a}M\ ,  \label{nlRicci}
\\
g_{a}^{c}{}n^{d}{}\widetilde{R}_{cd} &=&D_{c}K_{a}^{c}-D_{a}\left( K+%
\mathcal{K}\right) +\mathcal{K}_{a}L+\mathcal{L}_{\mathbf{l}}\mathcal{K}%
_{a}+M^{-1}K_{a}^{i}D_{i}M-M^{-1}\mathcal{K}D_{a}M\ ,  \label{Codazzi1} \\
g_{a}^{c}{}l^{d}{}\widetilde{R}_{cd} &=&D_{c}L_{a}^{c}-D_{a}\left( L-%
\mathcal{L}\right) +\mathcal{K}_{a}K+\mathcal{L}_{\mathbf{n}}\mathcal{K}%
_{a}+N^{-1}L_{a}^{i}D_{i}N+N^{-1}\mathcal{L}D_{a}N\ .  \label{Codazzi2}
\end{eqnarray}%
The last two Eqs. are the Codazzi equations. Note that 
$\mathcal{L}_{\mathbf{n}}K$ denotes the trace of 
$\mathcal{L}_{\mathbf{n}}K_{ab}$. 
The scalar curvatures are
related as 
\end{subequations}
\begin{eqnarray}
\widetilde{R} &=&R+K^{2}-3K_{ab}K^{ab}+2\mathcal{L}_{\mathbf{n}}\left( K+%
\mathcal{K}\right) -2N^{-1}D_{a}D^{a}N+2\mathcal{K}K+\mathcal{K}_{a}\mathcal{%
K}^{a}  \notag \\
&&-L^{2}+3L_{ab}L^{ab}-2\mathcal{L}_{\mathbf{l}}\left( L-\mathcal{L}\right)
-2M^{-1}D_{a}D^{a}M+2\mathcal{L}L-\left( NM\right) ^{-1}D_{a}ND^{a}M\ .
\end{eqnarray}%
By virtue of the (\ref{covardivs}) decompositions of covariant derivatives,
Eqs. (\ref{nnRicci}) and (\ref{llRicci}) can be put into the forms (\ref%
{Ricciprojdiv}) containing divergence terms. So can the scalar curvature.

Finally, the Einstein tensors are related as 
\begin{subequations}
\begin{eqnarray}
g_{a}^{c}{}g_{b}^{d}{}\widetilde{G}_{cd} &=&G_{ab}+K_{ab}\left( K+\mathcal{K}%
\right) -2K_{ac}K_{b}^{c}+\mathcal{L}_{\mathbf{n}}K_{ab}-N^{-1}D_{b}D_{a}N 
\notag \\
&&-L_{ab}\left( L-\mathcal{L}\right) +2L_{ac}L_{b}^{c}-\mathcal{L}_{\mathbf{l%
}}L_{ab}-M^{-1}D_{b}D_{a}M  \notag \\
&&-\frac{1}{2}g_{ab}\left[ K^{2}-3K_{cd}K^{cd}+\mathcal{L}_{\mathbf{n}%
}\left( K+\mathcal{K}\right) -2N^{-1}D_{c}D^{c}N\right]  \notag \\
&&-\frac{1}{2}g_{ab}\left[ -L^{2}+3L_{cd}L^{cd}-2\mathcal{L}_{\mathbf{l}%
}\left( L-\mathcal{L}\right) -2M^{-1}D_{c}D^{c}M\right]  \notag \\
&&-g_{ab}\left( \mathcal{K}K+\mathcal{K}_{c}\mathcal{K}^{c}-\left( NM\right)
^{-1}D_{c}ND^{c}M\right) \ ,  \label{Eingg} \\
n^{a}n^{b}\widetilde{G}_{ab} &=&\frac{1}{2}\left(
R+K^{2}-K_{ab}K^{ab}-L^{2}+3L_{ab}L^{ab}\right) +\mathcal{K}K-\mathcal{K}_{a}%
\mathcal{K}^{a}-\mathcal{K}^{2}  \notag \\
&&-\mathcal{L}_{\mathbf{l}}L+\mathcal{L}^{2}-M^{-1}D_{a}D^{a}M\ , \\
l^{a}l^{b}\widetilde{G}_{ab} &=&-\frac{1}{2}\left(
R-L^{2}+L_{ab}L^{ab}+K^{2}-3K_{ab}K^{ab}\right) -\mathcal{L}L+\mathcal{K}_{a}%
\mathcal{K}^{a}-\mathcal{L}^{2}  \notag \\
&&-\mathcal{L}_{\mathbf{n}}K+\mathcal{K}^{2}+N^{-1}D_{a}D^{a}N\ , \\
n^{a}l^{b}\widetilde{G}_{ab} &=&n^{a}l^{b}\widetilde{R}_{ab}\ , \\
g_{a}^{c}{}n^{d}{}\widetilde{G}_{cd} &=&g_{a}^{c}n^{d}\widetilde{R}_{cd}\ ,
\\
g_{a}^{c}{}l^{d}{}\widetilde{G}_{cd} &=&g_{a}^{c}l^{d}\widetilde{R}_{cd}\ .
\end{eqnarray}

By performing first the traditional $\left( s+1\right) +1$ decomposition and
then further splitting the $\left( s+1\right) $ dimensional spacelike
hypersurface into the $\ s$-dimensional brane and off-brane direction $%
\partial /\partial \chi $, we can give an independent proof of the relation
between the scalar curvatures. The twice contracted Gauss equation for the
hypersurface $\mathcal{S}_{t}$ of $\mathcal{B}$ is 
\end{subequations}
\begin{equation}
\widetilde{R}=\widehat{R}-\widehat{K}^{2}+\widehat{K}_{ab}\widehat{K}^{ab}-2%
\widetilde{\nabla }_{a}(\alpha ^{a}-\widehat{K}n^{a})\ .  \label{d3}
\end{equation}%
The extrinsic curvature $\widehat{K}_{ab}$ of $\mathcal{S}_{t}$ can be
further decomposed employing Eq. (\ref{Khatdecomp}), while for $\widehat{R}$
there is an other twice contracted Gauss equation, this time for the
hypersurface $\Sigma _{t,\chi }$ of $\mathcal{S}_{t}$: 
\begin{equation}
\widehat{R}=R+L^{2}-L_{ab}L^{ab}+2\widehat{D}_{a}(l^{c}\widehat{D}%
_{c}l^{a}-Ll^{a})\ .  \label{d4}
\end{equation}%
The latter relation and Eq. (\ref{Khatdecomp}) inserted in Eq. (\ref{d3})
and employing 
\begin{equation}
\widehat{D}_{a}(l^{c}\widehat{D}_{c}l^{a}-Ll^{a})=\widetilde{\nabla }%
_{a}(\lambda ^{a}-Ll^{a}-\mathcal{K}n^{a})-L\mathcal{L}-\lambda ^{a}\alpha
_{a}\ ,  \label{d5}
\end{equation}%
gives Eq. (\ref{curvscalar}) once again. 
\end{widetext}

\end{document}